# A New Non-MDS Hash Function Resisting Birthday Attack and Meet-in-the-middle Attack *

Shenghui Su [1, 2(✉)], Tao Xie [2], and Shuwang Lü [3, 4]

[1] Laboratory of Trusted Computing, Beijing University of Technology, Beijing 100124, PRC
[2] School of Computers, National University of Defense Technology, Changsha 410073, PRC
[3] School of Computers, University of Chinese Academy of Sciences, Beijing 100039, PRC
[4] Laboratory of Computational Complexity, BFID Corporation, Beijing 100098, PRC

**Abstract.** To examine the integrity and authenticity of an IP address efficiently and economically, this paper proposes a new non-Merkle-Damgård structural (non-MDS) hash function called JUNA that is based on a multivariate permutation problem and an anomalous subset product problem to which no subexponential time solutions are found so far. JUNA includes an initialization algorithm and a compression algorithm, and converts a short message of $n$ bits which is regarded as only one block into a digest of $m$ bits, where $80 \le m \le 232$ and $80 \le m \le n \le 4096$. The analysis and proof show that the new hash is one-way, weakly collision-free, and strongly collision-free, and its security against existent attacks such as birthday attack and meet-in-the-middle attack is to $O(2^m)$. Moreover, a detailed proof that the new hash function is resistant to the birthday attack is given. Compared with the Chaum-Heijst-Pfitzmann hash based on a discrete logarithm problem, the new hash is lightweight, and thus it opens a door to convenience for utilization of lightweight digital signing schemes.

## 1 Introduction

In recent years, the ECC-160 digital signing scheme, an analogue of the ElGamal digital signing scheme based on a discrete logarithm problem (DLP) in an elliptic curve group over a finite field, and some lightweight digital signing schemes have been utilized for RF (Radio Frequency) identity tags or non-RF identity tags. A RF identity tag contains an IC chip which is used to store signatures and other data, while a non-RF identity tag contains no IC chip because a signature by a lightweight or ultra-lightweight signing scheme may be converted into a short visual string less than 22 characters, and printed directly on a papery tag or label. In the near future, such non-RF tags will be applied to the identification, authentication, or anti-forgery of financial notes, bills, certificates, diplomas, and commodities, particularly including foods and drugs.

Additionally, message digests outputted by a hash function may be utilized to examine the integrity and authenticity of IP addresses in a transmitted data packet so as to prevent the source address and destination address from being tampered or forged.

It is well understood that we first need to extract the digest of a message by employing a hash function before signing the message [1][2][3]. Traditionally, a hash function consists of a compression function and the Merkle-Damgård iterative structure [4][5]. Let $\hat{h}$ be a hash function, and usually, it has the four properties as follows:

① given a message $\underline{m}$, it is very easy to calculate the message digest $\underline{d} = \hat{h}(\underline{m})$, where $\underline{d}$ is also called a hash output, namely $\hat{h}$ is computable;

② given a digest $\underline{d}$, it is very hard to calculate the message $\underline{m}$ according to $\underline{d} = \hat{h}(\underline{m})$, namely $\hat{h}$ is one-way;

③ given any arbitrary message $\underline{m}$, it is computationally infeasible to find another message $\underline{m}'$ such that $\hat{h}(\underline{m}) = \hat{h}(\underline{m}')$, namely $\hat{h}$ is weakly collision-free;

* This work is supported by MOST with Project 2009AA01Z441 and NSFC with Project 61472476. Email: reesse@126.com. Referring to: Theoretical Computer Science, v654, Nov 2016, pp.128–142.

④ it is computationally infeasible to find two arbitrary messages $\underline{m} \neq \underline{m}'$ such that $\hat{h}(\underline{m}) = \hat{h}(\underline{m}')$, namely $\hat{h}$ is strongly collision-free.

The word "infeasible" means that some problem cannot be solved at least in polynomial time or in tolerable subexponential time.

At present, SHA-1, SHA-256, and SHA-384 announced by NIST are among the hash functions that are believed to be secure though they each cannot resist birthday attack, which means that the security of each of them is nearly the $O(2^{m/2})$ magnitude, where $m$ is the bit-length of a message digest namely a hash output. It is well known that the output bit-lengths of these three functions are 160, 256, and 384 respectively.

When any of the three is practically paired with a lightweight signing scheme of which the modulus length is between 80 and 160 bits, its output must be adjusted to the range of the modulus length of the singing scheme with its security unchanged or corresponding to the signing scheme.

The modulus length of the optimized REESSE1+ signing scheme based on a transcendental logarithm problem and a polynomial root finding problem is 80 [6], and its security is the $2^{80}$ magnitude at present. When SHA-1 is paired with this signing scheme, the output of SHA-1 must be adjusted to 80 bits with its security unchanged. Again when SHA-256 is paired with ECC-160, the output of SHA-256 must be adjusted to 160 bits with its security being at least the $2^{80}$ magnitude.

Therefore, it is a problem in practice how to adjust a message digest from a classical hash function to the range of the modulus bit-length of a host signing scheme and to keep the security of the message digest being unchanged or corresponding to the host signing scheme.

In this paper, the authors devise a new non-Merkle-Damgård structural (non-MDS) hash function called JUNA which is based on a multivariate permutation problem (MPP) and an anomalous subset product problem (ASPP) [6][7], and includes two algorithms: an initialization algorithm and a compression algorithm, converts a short message or a message digest of $n$ bits into an output string of $m$ bits, where $80 \le m \le 232$ and $80 \le m \le n \le 4096$, and moreover ensures that the security of the output against existent collision attacks is to the $O(2^m)$ magnitude.

The new hash is efficient and economical in the integrity examination, and has two dominant novelties:

① devising the initialization algorithm based on a MPP which only has an exponential time solution currently, and makes the new hash function be able to resist birthday attack;

② devising the compression algorithm based on an ASPP which also only has an exponential time solution currently, and makes the new hash function be able to resist other conventional attacks, especially meet-in-the-middle attack.

The significance of the paper lies in the thing that a new non-iterative hash function with an $m$-bit output and the $O(2^m)$ magnitude security is first proposed by the authors while a classical iterative hash function with an $m$-bit output bears only the $O(2^{m/2})$ magnitude security.

Throughout the paper, unless otherwise specified, an even number $n \ge 80$ is the bit-length of a short message or the item-length of a sequence, the sign % denotes "modulo", $\bar{M}$ does "$M - 1$" with $M$ prime, $\lg x$ means a logarithm of $x$ to the base 2, $\neg b_i$ does NOT operation of a bit $b_i$, $\mathcal{P}$ does the maximal prime allowed in a coprime sequence, $|x|$ does the absolute value of a number $x$, $\|x\|$ does the order of $x$ % $M$, ¦$S$¦ does the size of a set $S$, and $\gcd(x, y)$ represents the greatest common divisor of two integers $x$ and $y$. Without ambiguity, "% $M$" is usually omitted in expressions.

## 2 Several Definitions

### 2.1 A Coprime Sequence

***Definition 1:*** If $A_1, \ldots, A_n$ are $n$ pairwise distinct positive integers such that $\forall\, A_i$ and $A_j$ $(i \ne j)$, either $\gcd(A_i, A_j) = 1$ or $\gcd(A_i, A_j) = F \ne 1$ with $(A_i / F) \nmid A_k$ and $(A_j / F) \nmid A_k$ $\forall\, k\ (\ne i, j) \in [1, n]$, these ordered integers are called a coprime sequence, denoted by $\{A_1, \ldots, A_n\}$, and shortly $\{A_i\}$.

Notice that the elements of a coprime sequence are not necessarily pairwise coprime, but a sequence of which all the elements are pairwise coprime is a coprime sequence.

For example, {15, 29, 163, 31, 37, 509, 21, 1669}, {37, 23, 7, 1009, 3, 1999, 937, 17}, {3607, 61, 59, 97, 1021, 211, 863, 2039}, and {10, 211, 127, 3, 14, 1021, 2017, 263} are four coprime sequences separately.

***Property 1:*** Let $\{A_1, \ldots, A_n\}$ be a coprime sequence. If randomly select $k \in [1, n]$ elements $A_{x_1}, \ldots, A_{x_k}$ from the sequence, then the mapping from a subset $\{A_{x_1}, \ldots, A_{x_k}\}$ to a subset product $G = \prod_{i=1}^{k} A_{x_i}$ is one-to-one, namely the mapping from $b_1 \ldots b_n$ to $G = \prod_{i=1}^{n} A_i^{b_i}$ is one-to-one, where $b_1 \ldots b_n$ is a bit string.

Refer to [6] for its proof.

## 2.2 A Bit Shadow and a Bit Long-shadow

***Definition 2:*** Let $b_1 \ldots b_n \neq 0$ be a bit string. Then $\underset{.}{b}_i$ with $i \in [1, n]$ is called a bit shadow if it comes from such a rule:

① $\underset{.}{b}_i = 0$ if $b_i = 0$;

② $\underset{.}{b}_i = 1 +$ the number of successive 0-bits before $b_i$ if $b_i = 1$; or

③ $\underset{.}{b}_i = 1 +$ the number of successive 0-bits before $b_i$ + the number of successive 0-bits after the rightmost 1-bit if $b_i$ is the leftmost 1-bit.

Notice that the third point of this definition is slightly different from that in [6].

For example, let $b_1 \ldots b_8 = 01010100$, then $\underset{.}{b}_1 \ldots \underset{.}{b}_8 = 04020200$.

***Fact 1:*** Let $\underset{.}{b}_1 \ldots \underset{.}{b}_n$ be the bit shadow string of $b_1 \ldots b_n \neq 0$. Then there is $\sum_{i=1}^{n} \underset{.}{b}_i = n$.

*Proof:*

According to Definition 2, every bit of $b_1 \ldots b_n$ is considered into $\sum_{i=1}^{k} \underset{.}{b}_{x_i}$, where $\underset{.}{b}_{x_1}, \ldots, \underset{.}{b}_{x_k}$ are 1-bit shadows in the string $\underset{.}{b}_1 \ldots \underset{.}{b}_n$, and there is $\sum_{i=1}^{k} \underset{.}{b}_{x_i} = n$.

On the other hand, there is $\sum_{j=1}^{n-k} \underset{.}{b}_{y_j} = 0$, where $\underset{.}{b}_{y_1}, \ldots, \underset{.}{b}_{y_{n-k}}$ are 0-bit shadows.

In total, there is $\sum_{i=1}^{n} \underset{.}{b}_i = n$. □

***Property 2:*** Let $\{A_1, \ldots, A_n\}$ be a coprime sequence, and $\underset{.}{b}_1 \ldots \underset{.}{b}_n$ be the bit shadow string of $b_1 \ldots b_n \neq 0$. Then the mapping from $b_1 \ldots b_n$ to $G = \prod_{i=1}^{n} A_i^{\underset{.}{b}_i}$ is one-to-one.

*Proof:*

Step 1. Let $b_1 \ldots b_n$ and $b'_1 \ldots b'_n$ be two different nonzero bit strings, and $\underset{.}{b}_1 \ldots \underset{.}{b}_n$ and $\underset{.}{b}'_1 \ldots \underset{.}{b}'_n$ be the two corresponding bit shadow strings.

If $\underset{.}{b}_1 \ldots \underset{.}{b}_n = \underset{.}{b}'_1 \ldots \underset{.}{b}'_n$, then by Definition 2, there is $b_1 \ldots b_n = b'_1 \ldots b'_n$.

In addition, for any arbitrary bit shadow string $\underset{.}{b}_1 \ldots \underset{.}{b}_n$, there always exists a preimage $b_1 \ldots b_n$. Thus, the mapping from $b_1 \ldots b_n$ to $\underset{.}{b}_1 \ldots \underset{.}{b}_n$ is one-to-one.

Step 2. Obviously the mapping from $\underset{.}{b}_1 \ldots \underset{.}{b}_n$ to $\prod_{i=1}^{n} A_i^{\underset{.}{b}_i}$ is surjective.

Again presuppose that $\prod_{i=1}^{n} A_i^{\underset{.}{b}_i} = \prod_{i=1}^{n} A_i^{\underset{.}{b}'_i}$ for $\underset{.}{b}_1 \ldots \underset{.}{b}_n \neq \underset{.}{b}'_1 \ldots \underset{.}{b}'_n$.

Since $\{A_1, \ldots, A_n\}$ is a coprime sequence, and $A_i^{\underset{.}{b}_i}$ either equals 1 with $\underset{.}{b}_i = 0$ or contains the same prime factors as those of $A_i$ with $\underset{.}{b}_i \neq 0$, we can obtain $\underset{.}{b}_1 \ldots \underset{.}{b}_n = \underset{.}{b}'_1 \ldots \underset{.}{b}'_n$ from $\prod_{i=1}^{n} A_i^{\underset{.}{b}_i} = \prod_{i=1}^{n} A_i^{\underset{.}{b}'_i}$, which is in direct contradiction to $\underset{.}{b}_1 \ldots \underset{.}{b}_n \neq \underset{.}{b}'_1 \ldots \underset{.}{b}'_n$.

Therefore, the mapping from $\underset{.}{b}_1 \ldots \underset{.}{b}_n$ to $\prod_{i=1}^{n} A_i^{\underset{.}{b}_i}$ is injective [8].

In summary, the mapping from $\underset{.}{b}_1 \ldots \underset{.}{b}_n$ to $\prod_{i=1}^{n} A_i^{\underset{.}{b}_i}$ is one-to-one, and further the mapping from $b_1 \ldots b_n$ to $\prod_{i=1}^{n} A_i^{\underset{.}{b}_i}$ is also one-to-one. □

***Definition 3:*** Let $\underset{.}{b}_1 \ldots \underset{.}{b}_n$ be the bit shadow string of $b_1 \ldots b_n \neq 0$. Then $\bar{b}_i = \underset{.}{b}_i 2^{\partial_i}$ with $i \in [1, n]$ is called a bit long-shadow, where $\partial_i = b_{i+(-1)^{\lfloor 2(i-1)/n \rfloor}(n/2)} = 0$ or 1.

According to Definition 3, it is not difficult to understand that for every $\bar{b}_i$, there is $0 \le \bar{b}_i \le n$ when $b_1 \ldots b_n \neq 0$.

For example, let $b_1 \ldots b_8 = 01010100$, then $\bar{b}_1 \ldots \bar{b}_8 = 08020400$.

***Fact 2:*** Let $\bar{b}_1 \ldots \bar{b}_n$ be the bit long-shadow string of $b_1 \ldots b_n \neq 0$. Then there is $n \le \sum_{i=1}^{n} \bar{b}_i \le 2n$.

*Proof:*

By Definition 3 and Fact 1, we have

$$\sum_{i=1}^{n} \bar{b}_i = \sum_{i=1}^{n} \underset{.}{b}_i 2^{\partial_i} \text{ and } \sum_{i=1}^{n} \underset{.}{b}_i = n.$$

If every $b_i = 1$, namely every $\partial_i = 1$, then

$$\sum_{i=1}^{n} \bar{b}_i = \sum_{i=1}^{n} \underset{.}{b}_i 2^{\partial_i} = 2\sum_{i=1}^{n} \underset{.}{b}_i = 2n.$$

Again, by Definition 3, not all the bits of $b_1 \ldots b_n$ are zero.

If there exists only one nonzero bit in $b_1 \ldots b_n$ — $b_x = 1$ with $x \in [1, n]$ for example, then

$$\sum_{i=1}^{n} \bar{b}_i = \sum_{i=1}^{n} \underset{.}{b}_i 2^{\partial_i} = \underset{.}{b}_x 2^{\partial_x} = \underset{.}{b}_x = n,$$

where $\partial_x = b_{x+(-1)^{\lfloor 2(x-1)/n \rfloor}(n/2)} = 0$ due to $b_x$ being the unique nonzero bit.

Thus, it holds that $n \le \sum_{i=1}^{n} \bar{b}_i \le 2n$. □

***Property 3:*** Let $\bar{b}_1 \ldots \bar{b}_n$ be the bit long-shadow string of $b_1 \ldots b_n \ne 0$. Then the mapping from $b_1 \ldots b_n$ to $\bar{b}_1 \ldots \bar{b}_n$ is one-to-one.

*Proof:*

On one hand, assume that a bit string $b_1 \ldots b_n \ne 0$ is known.

It is understood from Definition 3 that $\bar{b}_i = \underset{.}{b}_i 2^{\partial_i}$ for each $i$.

Because when $b_1 \ldots b_n$ is known, $\underset{.}{b}_1 \ldots \underset{.}{b}_n$ and $\partial_1 \ldots \partial_n$ can be respectively determined, $\bar{b}_1 \ldots \bar{b}_n$ can also be determined uniquely.

On the other hand, assume that a bit long-shadow string $\bar{b}_1 \ldots \bar{b}_n$ is known.

According to $\bar{b}_i = \underset{.}{b}_i 2^{\partial_i}$ and $\bar{b}_i = 0$ with $\underset{.}{b}_i = 0$, where $\partial_i = b_{i+(-1)^{\lfloor 2(i-1)/n \rfloor}(n/2)}$, we can determinate $b_i$ for $i$ = 1, …, $n$ as follows.

① Case of $\bar{b}_i = 0$

If $\bar{b}_i = 0$, then $\underset{.}{b}_i = 0$, and set $b_i = 0$.

② Case of $\bar{b}_i \ne 0$

If $\bar{b}_i \ne 0$, then $\underset{.}{b}_i \ne 0$, and set $b_i = 1$.

In this way, the value of every $b_i$ can be determined uniquely.

In summary, the mapping from $b_1 \ldots b_n$ to $\bar{b}_1 \ldots \bar{b}_n$ is one-to-one. □

## 2.3 A Lever Function

The devising of the initialization algorithm of the new hash function is based on the intractable problem $C_i \equiv (A_i W^{\ell(i)})^{\delta}$ (% $M$) for $i$ = 1, …, $n$ which is first utilized for the REESSE1+ asymmetric cryptosystem, where the exponent $\ell(i)$ is called a lever function [6].

***Definition 4:*** The secret parameter $\ell(i)$ in the transform of a non-iterative hash function is called a lever function, if it has the following features:

① $\ell(.)$ is an injection from the domain $\{1, \ldots, n\}$ to the codomain $\Omega \subset \{5, \ldots, \bar{M}\}$ with $\bar{M}$ large;

② the mapping between $i$ and $\ell(i)$ is established randomly without an analytical expression;

③ an attacker has to be faced with all the permutations of elements in $\Omega$ when inferring a related private parameter from a public parameter or an initial value;

④ the owner of the private parameter only need to consider the polynomial arithmetic of elements in $\Omega$ when decrypting a ciphertext or seeking a collision.

Feature ③ and ④ make it clear that if $n$ is large enough, it is infeasible for the attacker to search all the permutations of elements in $\Omega$ exhaustively while the decryption or collision computation by the owner of the private parameter is feasible. Thus, the amount of calculation on $\ell(.)$ is large at "a public terminal", and is small at "a private terminal".

***Property 4* (Indeterminacy of $\ell(.)$):** Let $\delta = 1$ and $C_i \equiv (A_i W^{\ell(i)})^{\delta}$ (% $M$) with $\ell(i) \in \Omega = \{5, \ldots, n + 4\}$ and $A_i \in \Lambda = \{2, \ldots, \bar{P} \mid 863 \le \bar{P} \le 1201\}$ for $i$ = 1, …, $n$. Then $\forall W (\|W\| \ne \bar{M}) \in (1, \bar{M})$, and $\forall x, y, z$ $(x \ne y \ne z) \in [1, n]$,

① when $\ell(x) + \ell(y) = \ell(z)$, there is $\ell(x) + \|W\| + \ell(y) + \|W\| \ne \ell(z) + \|W\|$ (% $\bar{M}$);

② when $\ell(x) + \ell(y) \ne \ell(z)$, there always exist

$$C_x \equiv A'_x W'^{\ell'(x)} \text{ (\% } M\text{)},\ C_y \equiv A'_y W'^{\ell'(y)} \text{ (\% } M\text{), and } C_z \equiv A'_z W'^{\ell'(z)} \text{ (\% } M\text{)}$$

such that $\ell'(x) + \ell'(y) \equiv \ell'(z)$ (% $\bar{M}$) with the constraint $A'_z \le \bar{P}$.

*Proof:*

① It is easy to understand that

$$W^{\ell(x)} \equiv W^{\ell(x)+\|W\|},\ W^{\ell(y)} \equiv W^{\ell(y)+\|W\|}, \text{ and}$$
$$W^{\ell(z)} \equiv W^{\ell(z)+\|W\|} \text{ (\% } M\text{)}.$$

Due to $\|W\| \ne \bar{M}$, $2\|W\| \ne \|W\|$, and $\ell(x) + \ell(y) = \ell(z)$, it follows that

$$\ell(x) + \|W\| + \ell(y) + \|W\| \ne \ell(z) + \|W\| \text{ (\% } \bar{M}\text{)}.$$

However, it should be noted that when $\|W\| = \bar{M}$, there is $\ell(x) + \|W\| + \ell(y) + \|W\| \equiv \ell(z) + \|W\|$ (% $\bar{M}$).

② Let $\bar{O}_d$ be an oracle on solving a discrete logarithm problem.

Suppose that $W' \in [1, \bar{M}]$ is a generator of $(\mathbb{Z}_M^*, \cdot)$.

In light of group theories, $\forall A'_z \in \{2, \ldots, \bar{P}\}$, the congruence

$$C_z \equiv A'_z W'^{\ell'(z)} \text{ (\% } M)$$

has a solution. Then, $\ell'(z)$ may be taken through $\bar{O}_d$.

$\forall\ \ell'(x) \in [1, \bar{M}]$, and let

$$\ell'(y) \equiv \ell'(z) - \ell'(x) \text{ (\% } \bar{M}).$$

Further, from the congruences $C_x \equiv A'_x W'^{\ell'(x)}$ (% $M$) and $C_y \equiv A'_y W'^{\ell'(y)}$ (% $M$), we can obtain many distinct pairs $(A'_x, A'_y)$, where $A'_x, A'_y \in (1, M)$, and $\ell'(x) + \ell'(y) \equiv \ell'(z)$ (% $\bar{M}$).

In this way, Property 4 is proven. □

Notice that letting $\Omega = \{5, \ldots, n + 4\}$, namely every $\ell(i) \ge 5$ makes seeking $W$ from $W^{\ell(i)} \equiv A_i^{-1} C_i$ (% $M$) face an unsolvable Galois group when the value of $A_i \le \bar{P}$ is guessed [9], and moreover Property 4 still holds when $\Omega$ is any subset containing $n$ elements from $\{1, \ldots, \bar{M}\}$.

Property 4 manifests that will continued fraction attack on $C_i \equiv A_i W^{\ell(i)}$ (% $M$) by Theorem 12.19 in Section 12.3 of [10] be utterly ineffectual only if elements in $\Omega$ are fitly selected [11].

# 3 Design of the New Non-iterative Hash Function

The Chaum-Heijst-Pfitzmann hash function, a non-iterative one, is appreciable. It is based on a discrete logarithm problem, and proved to be strongly collision-free [12].

The new non-iterative hash function is composed of two algorithms which contain two main parameters $m$ and $n$, where $m$ denotes the bit-length of a modulus utilized in the new hash, $n$ denotes the bit-length of a short message or a message digest from a classical hash function, and there are $80 \le m \le 232$ with $80 \le m \le n \le 4096$.

Additionally, $\Lambda$ and $\Omega$ are two integral sets. Their lengths are selected as $2^{10} \le |\Lambda| \le 2^{32}$ and $n \le |\Omega| = \tilde{n} \le 2^{32}$, and moreover make $2n^5|\Omega||\Lambda|^5 \ge 2^m$ (see Section 4.1.1). Notice that $2^{10} \le |\Lambda| \le 2^{32}$ means $10 \le \lceil \lg \bar{P} \rceil \le 32$.

For example, as $m = 80 \le n$, there should be $|\Lambda| = 2^{10}$ and $|\Omega| = n$; as $m = 96 \le n$, should $|\Lambda| = 2^{12}$ and $|\Omega| = n$; as $m = 112 \le n$, should $|\Lambda| = 2^{14}$ and $|\Omega| = n$; as $m = 128 \le n$, should $|\Lambda| = 2^{16}$ and $|\Omega| = 2^{12}$; as $m = 232 \le n$, should $|\Lambda| = 2^{32}$ and $|\Omega| = 2^{32}$.

## 3.1 Initialization Algorithm

This algorithm is employed by an authoritative third party or the owner of a key pair, and only needs to be executed one time.

INPUT: the bit-length $m$ of a modulus with $80 \le m \le 232$;
the item-length $n$ of a sequence with $80 \le m \le n \le 4096$;
the maximal prime $\bar{P}$ with $10 \le \lceil \lg \bar{P} \rceil \le 32$;
the size $\tilde{n}$ of the set $\Omega$ with $2\tilde{n}n^5\bar{P}^5 \ge 2^m$ and $n \le \tilde{n} \le 2^{32}$.

S1: Produce $\Lambda \leftarrow \{2, 3, \ldots, \bar{P}\}$;
produce a random coprime sequence $\{A_1, \ldots, A_n \mid A_i \in \Lambda\}$.

S2: Find a prime $M$ with $\lceil \lg M \rceil = m$ such that $\bar{M} / 2$ is a prime,
or the least prime factor of $\bar{M} / 2 > 4n(2\tilde{n} + 3)$.

S3: Pick $W \in (1, \bar{M})$ making $\|W\| \ge 2^{m - \lceil \lg \bar{P} \rceil}$;
pick $\delta \in (1, \bar{M})$ making $\gcd(\delta, \bar{M}) = 1$.

S4: Randomly yield $\Omega \leftarrow \{+/-5, +/-7, \ldots, +/-(2\tilde{n} + 3)\}$;
randomly select pairwise distinct $\ell(i) \in \Omega$ for $i = 1, \ldots, n$.

S5: Compute $C_i \leftarrow (A_i W^{\ell(i)})^\delta$ % $M$ for $i = 1, \ldots, n$.

OUTPUT: an initial value $(\{C_i\}, M)$ which is public to the people.

A private parameter $(\{A_i\}, \{\ell(i)\}, W, \delta)$ may be discarded, but must not be divulged.

At S3, to seek $W$, let $W \equiv g^{\bar{M}/F}$ (% $M$), where $g$ is a generator of $(\mathbb{Z}_M^*, \cdot)$ obtained through Algorithm 4.80 in Section 4.6 of [1], and $F < 2^{\lceil \lg \bar{P} \rceil}$ is a factor of $\bar{M}$.

At S4, $\Omega = \{+/-5, +/-7, \ldots, +/-(2\tilde{n} + 3)\}$ indicates that $\Omega$ is one of $2^{\tilde{n}}$ potential sets, indeterminate, and unknown to the public, where "+/–" means the selection of the "+" or "–" sign. Notice that in the arithmetic modulo $\bar{M}$, $-x$ represents $\bar{M} - x$.

***Definition 5:*** Given $(\{C_i\}, M)$, seeking the original $(\{A_i\}, \{\ell(i)\}, W, \delta)$ from $C_i \equiv (A_i W^{\ell(i)})^\delta$ (% $M$) with $A_i \in \{2, 3, \dots, \grave{P} \mid 10 \le \lceil \lg \grave{P} \rceil \le 32\}$ and $\ell(i) \in \{+/-5, +/-7, \dots, +/-(2\tilde{n} + 3) \mid n \le \tilde{n} \le 2^{32}\}$ for $i = 1, \dots, n$ is referred to as a multivariate permutation problem, shortly MPP [6].

***Property 5:*** The MPP $C_i \equiv (A_i W^{\ell(i)})^\delta$ (% $M$) with $A_i \in \{2, 3, \dots, \grave{P} \mid 10 \le \lceil \lg \grave{P} \rceil \le 32\}$ and $\ell(i) \in \{+/-5, +/-7, \dots, +/-(2\tilde{n} + 3) \mid n \le \tilde{n} \le 2^{32}\}$ for $i = 1, \dots, n$ is computationally at least equivalent to the discrete logarithm problem (DLP) in the same prime field.

### 3.2 Compression Algorithm

This algorithm is employed by one who wants to obtain a short message digest.

INPUT: an initial value $(\{C_1, \dots, C_n\}, M)$, where $\lceil \lg M \rceil = m$ with $80 \le m \le n \le 4096$;

A short message (or a digest from a classical hash function) $b_1 \dots b_n \ne 0$.

S1: Set $k \leftarrow 0$, $i \leftarrow 1$.

S2: If $b_i = 0$ then

S2.1: let $k \leftarrow k + 1$, $\underset{\cdot}{b}_i \leftarrow 0$

else

S2.2: if $i = k + 1$ then let $\bar{s} \leftarrow i$;

S2.3: let $\underset{\cdot}{b}_i \leftarrow k + 1$, $k \leftarrow 0$.

S3: Let $i \leftarrow i + 1$;

if $i \le n$ then go to S2.

S4: Compute $\underset{\cdot}{b}_{\bar{s}} \leftarrow \underset{\cdot}{b}_{\bar{s}} + k$.

S5: Compute $\underline{d} \leftarrow \prod_{i=1}^{n} C_i^{\bar{b}_i} \% M$,

where $\bar{b}_i = \underset{\cdot}{b}_i 2^{\partial_i}$ with $\partial_i = b_{i + (-1)^{\lfloor 2(i-1)/n \rfloor}(n/2)}$.

OUTPUT: a digest $\underline{d} \equiv \prod_{i=1}^{n} C_i^{\bar{b}_i}$ (% $M$) of which the bit-length is $m$.

It is easily known from Definition 3 that the max of $\{\bar{b}_1, \dots, \bar{b}_n\}$ is less than or equal to $n$ when $b_1 \dots b_n \ne 0$.

***Definition 6:*** Given $(\underline{d}, M)$, seeking the original $\bar{b}_1 \dots \bar{b}_n$ from $\underline{d} \equiv \prod_{i=1}^{n} C_i^{\bar{b}_i}$ (% $M$), where $\bar{b}_i = \underset{\cdot}{b}_i 2^{\partial_i}$ with $\partial_i = b_{i + (-1)^{\lfloor 2(i-1)/n \rfloor}(n/2)}$ and $\underset{\cdot}{b}_i$ being a bit shadow is referred to as an anomalous subset product problem, shortly ASPP [6].

***Property 6:*** The ASPP $\underline{d} \equiv \prod_{i=1}^{n} C_i^{\bar{b}_i}$ (% $M$), where $\bar{b}_i = \underset{\cdot}{b}_i 2^{\partial_i}$ with $\partial_i = b_{i + (-1)^{\lfloor 2(i-1)/n \rfloor}(n/2)}$ and $\underset{\cdot}{b}_i$ being a bit shadow is computationally at least equivalent to the DLP in the same prime field.

### 3.3 Proofs of Property 5 and 6

***Definition 7:*** Let $A$ and $B$ be two computational problems. $A$ is said to reduce to $B$ in polynomial time, written as $A \le_T^P B$, if there is an algorithm for solving $A$ which calls, as a subroutine, a hypothetical algorithm for solving $B$, and runs in polynomial time, excluding the time of the algorithm for solving $B$ [1][13].

The hypothetical algorithm for solving $B$ is called an oracle. It is easy to understand that no matter what the time complexity of the oracle is, it does not influence the result of the comparison.

$A \le_T^P B$ means that the difficulty of $A$ is not greater than that of $B$, namely the time complexity of the fastest algorithm for solving $A$ is not greater than that of the fastest algorithm for solving $B$ when all polynomial times are treated as the identical magnitude. Concretely speaking, if $A$ cannot be solved in polynomial or subexponential time, correspondingly $B$ cannot also be solved in polynomial or subexponential time; and if $B$ can be solved in polynomial or subexponential time, correspondingly $A$ can also be solved in polynomial or subexponential time.

***Definition 8:*** Let $A$ and $B$ be two computational problems. If $A \le_T^P B$ and $B \le_T^P A$, then $A$ and $B$ are said to be computationally equivalent, written as $A =_T^P B$ [1][13].

$A =_T^P B$ means that either if $A$ is a intractability with a certain complexity on a condition that its dominant variable approaches a large number, $B$ is also a intractability with the same complexity on the identical condition; or both $A$ and $B$ can be solved in linear or polynomial time.

Obviously, Definition 7 and 8 gives a partial order relation among the complexities or difficulties of computational problems [14], and suggest a reductive proof method called polynomial time Turing reduction (PTR) [13].

In addition, for convenience sake, let $\hat{H}(y = f(x))$ represent the complexity or difficulty of the problem of solving $y = f(x)$ for $x$ [15].

What follows is the proof of Property 5.

*Proof:*

Firstly, we systematically consider $C_i \equiv (A_i W^{\ell(i)})^\delta$ (% $M$) for $i = 1, \dots, n$.

Assume that each $g_i \equiv A_i W^{\ell(i)}$ (% $M$) with $\ell(i) \in \{+/-5, +/-7, \dots, +/-(2\tilde{n} + 3) \mid n \le \tilde{n} \le 2^{32}\}$ is a constant.

Let

$$g_i \equiv g^{x_i} \ (\%\ M), \text{ and } z_i \equiv \delta x_i \ (\%\ \bar{M}),$$

where $g \in \mathbb{Z}_M^*$ be a generator.

Then, there is

$$C_i \equiv g_i^\delta \equiv g^{\delta x_i} \ (\%\ M) \text{ for } i = 1, \dots, n.$$

Again let

$$\delta x_i \equiv z_i \ (\%\ \bar{M}).$$

Further

$$C_i \equiv g^{z_i} \ (\%\ M) \text{ for } i = 1, \dots, n.$$

The above expression corresponds to the fact that in the ElGamal cryptosystem where many users share the modulus and a key generator, User 1 acquires a private key $z_1$ and a public key $C_1$, …, and User $n$ acquires a private key $z_n$ and a public key $C_n$. It is well known that in this case, the attack of an adversary is still faced with the DLP, namely seeking $z_i$ from the simultaneous equation $C_i \equiv g^{z_i}$ (% $M$) for $i = 1, \dots, n$ is computationally equivalent to the DLP [1].

Thus, when every $g_i$ is weakened to a constant, seeking $\delta$ from $C_i \equiv g_i^\delta$ (% $M$) for $i = 1, \dots, n$ is computationally equivalent to the DLP, which indicates that when every $g_i$ is not a constant, seeking $g_i$ and $\delta$ from $C_i \equiv g_i^\delta$ (% $M$) for $i = 1, \dots, n$ is computationally at least equivalent to the DLP.

Secondly, singly consider a certain $C_i$, where the subscript $i$ is designated.

Assume that $\bar{O}_m(C_i, M, \underline{R})$ is an oracle on solving $C_i \equiv g_i^\delta$ (% $M$) for $g_i$ and $\delta$, where $i$ is in $\{1, \dots, n\}$, and $\underline{R}$ is a constraint on $g_i$ such that the original $g_i$ and $\delta$ can be found.

Let $y \equiv g^x$ (% $M$) be of the DLP. Then, by calling $\bar{O}_m(y, M, g)$, $x$ can be obtained.

According to Definition 7, there is

$$\hat{H}(y \equiv g^x \ (\%\ M)) \le_T^P \hat{H}(C_i \equiv g_i^\delta \ (\%\ M)),$$

which indicates that when only a certain $g_i$ is known, seeking $g_i$ and $\delta$ from $C_i \equiv g_i^\delta$ (% $M$) is computationally at least equivalent to the DLP.

Integrally, we say that seeking the original $\{A_i\}$, $\{\ell(i)\}$, $W$, and $\delta$ from the public key $C_i \equiv (A_i W^{\ell(i)})^\delta$ (% $M$) for $i = 1, \dots, n$ is computationally at least equivalent to the DLP in the same prime field. □

What follows is the proof of Property 6.

*Proof:*

Assume that $\bar{O}_a(\underline{d}, C_1, \dots, C_n, M)$ is an oracle on solving $\underline{d} \equiv \prod_{i=1}^n C_i^{\bar{b}_i}$ (% $M$) for $\bar{b}_1 \dots \bar{b}_n$, where $\bar{b}_1 \dots \bar{b}_n$ is the bit long-shadow string of $b_1 \dots b_n$.

Particularly, when $C_1 = \dots = C_n = C$, define

$$\underline{d} \equiv \prod_{i=1}^n C^{(n+1)^{n-i}\bar{b}_i} \equiv \prod_{i=1}^n (C^{(n+1)^{n-i}})^{\bar{b}_i} \ (\%\ M)$$

with $0 \le \bar{b}_i \le n$, and define the corresponding oracle as $\bar{O}_a(\underline{d}, C^{(n+1)^{n-1}}, \dots, C^{(n+1)^0}, M)$.

Let $\bar{G}_1 \equiv \prod_{i=1}^n C_i^{b_i}$ (% $M$) be of the subset product problem (SPP) [6][7][16].

Since there is $0 \le b_i \le \bar{b}_i$, and the mapping from $\bar{b}_1 \dots \bar{b}_n$ to $b_1 \dots b_n$ is one-to-one, by calling $\bar{O}_a(\bar{G}_1, C_1, \dots, C_n, M)$, we can find $b_1 \dots b_n$.

By Definition 7, there is

$$\hat{H}(\bar{G}_1 \equiv \prod_{i=1}^n C_i^{b_i} \ (\%\ M)) \le_T^P \hat{H}(\underline{d} \equiv \prod_{i=1}^n C_i^{\bar{b}_i} \ (\%\ M)).$$

By Property 5 in [6], there is

$$\hat{H}(y \equiv g^x \ (\%\ M)) \le_T^P \hat{H}(\bar{G}_1 \equiv \prod_{i=1}^n C_i^{b_i} \ (\%\ M)).$$

Further, by transitivity, there is

$$\hat{H}(y \equiv g^x \ (\%\ M)) \le_T^P \hat{H}(\underline{d} \equiv \prod_{i=1}^n C_i^{\bar{b}_i} \ (\%\ M)).$$

Therefore, solving $\underline{d} \equiv \prod_{i=1}^n C_i^{\bar{b}_i}$ (% $M$) for $\bar{b}_1 \dots \bar{b}_n$ is at least equivalent to the DLP in the same prime field in computational complexity. □

# 4 Security Analysis of the New Hash Function

It is should be noted that $\lceil \lg M \rceil = m$, but not $n$, is the security dominant parameter of the new non-iterative hash function.

## 4.1 Security of the Initialization Algorithm

Clearly, the security of the initialization algorithm depends on the security of the MPP $C_i \equiv (A_i W^{\ell(i)})^{\delta}$ (% $M$) with $A_i \in \Lambda = \{2, 3, \ldots, Ƥ \mid 10 \le \lceil \lg Ƥ \rceil \le 32\}$ and $\ell(i) \in \Omega = \{+/-5, +/-7, \ldots, +/-(2\tilde{n} + 3) \mid n \le \tilde{n} \le 2^{32}\}$ for $i = 1, \ldots, n$.

In [6], we analyze the security of the MPP $C_i \equiv (A_i W^{\ell(i)})^{\delta}$ (% $M$) with $A_i \in \{2, 3, \ldots, Ƥ \mid 863 \le Ƥ \le 1201\}$ and $\ell(i) \in \{5, 7, \ldots, (2n + 3)\}$ for $i = 1, \ldots, n$ from the three aspects, discover no subexponential time solution to it, and contrarily, find some evidence which inclines people to believe that the MPP is computationally harder than the DLP.

Considering that the set $\Omega$ is different from the old in [6], and the range of $Ƥ$ is larger than the old in [6], we will analyze the security of the MPP with the different restrictions additionally.

### 4.1.1 Ineffectualness of Presupposing $\ell(x_1) + \ell(x_2) = \ell(y_1) + \ell(y_2)$

Because of $\Omega = \{+/-5, +/-7, \ldots, +/-(2\tilde{n} + 3)\}$, when the absolute values $|\ell(x_1)|$, $|\ell(x_2)|$, $|\ell(y_1)|$, $|\ell(y_2)|$ are determined, the value $\ell(x_1) + \ell(x_2) - (\ell(y_1) + \ell(y_2))$ has $2^4 = 16$ possible cases, which enhances the indeterminacy of the lever function, and increases the complexity of an attack task for cracking the MPP to some extent.

Adversaries may try to eliminate $W$ through judging $\ell(x_1) + \ell(x_2) = \ell(y_1) + \ell(y_2)$.

$\forall\, x_1, x_2, y_1, y_2 \in [1, n]$, presuppose that $\ell(x_1) + \ell(x_2) = \ell(y_1) + \ell(y_2)$ holds.

Let

$$G_z \equiv C_{x_1} C_{x_2} (C_{y_1} C_{y_2})^{-1} \text{ (\% } M\text{), namely}$$

$$G_z \equiv (A_{x_1} A_{x_2} (A_{y_1} A_{y_2})^{-1})^{\delta} \text{ (\% } M\text{)}.$$

If the adversaries divine the values of $A_{x_1}$, $A_{x_2}$, $A_{y_1}$, $A_{y_2}$, and compute $u$, $v_{x_1}$, $v_{x_2}$, $v_{y_1}$, $v_{y_2}$ in at least $L_M[1/3, 1.923]$ time such that

$$G_z \equiv g^u,\ A_{x_1} \equiv g^{v_{x_1}},\ A_{x_2} \equiv g^{v_{x_2}},\ A_{y_1} \equiv g^{v_{y_1}},\ A_{y_2} \equiv g^{v_{y_2}} \text{ (\% } M\text{)},$$

where $g$ is a generator of $(\mathbb{Z}_M^*, \cdot)$, then

$$u \equiv (v_{x_1} + v_{x_2} - v_{y_1} - v_{y_2})\delta \text{ (\% } \bar{M}\text{)}.$$

If $\gcd(v_{x_1} + v_{x_2} - v_{y_1} - v_{y_2}, \bar{M}) \mid u$, the congruence in $\delta$ has solutions. Because each of $A_{x_1}$, $A_{x_2}$, $A_{y_1}$, $A_{y_2}$ may traverse the interval $\Lambda$, and the subscripts $x_1, x_2, y_1, y_2$ are unfixed, the number of potential values of $\delta$ is about $n^4 |\Lambda|^4$. Notice that the number of non-repeated values of $\delta$ will be less than $2^m$.

In succession, we need to seek $W$.

Now, the most effectual approach to seeking $W$ is that for every $i$, the adversaries fix a value of $\delta$, divine $A_i$ and $\ell(i)$, and find the set $\bar{V}_i$ according to $C_i \equiv (A_i W^{\ell(i)})^{\delta}$ (% $M$), where $\bar{V}_i$ is the set of possible values of $W$ meeting $C_i \equiv (A_i W^{\ell(i)})^{\delta}$ (% $M$) for $i = 1, \ldots, n$. If there exist $W_1 \in \bar{V}_1, \ldots, W_n \in \bar{V}_n$ which are pairwise equal, the divination of $\delta$, $\{A_i\}$, and $\{\ell(i)\}$ is thought right; else fix another value of $\delta$, repeat the above process.

Notice that due to $\bar{M} / 2$ = a prime or the least prime factor of $\bar{M} / 2 > 4n(2\tilde{n} + 3)$, $W^{\ell(i)} \equiv C_i^{\delta^{-1}} A_i^{-1}$ (% $M$) can be solved in polynomial time, and besides letting $W = g^u$ % $M$ is unnecessary.

It is not difficulty to understand that the size of every $\bar{V}_i$ is about $(2|\Omega|)|\Lambda|$.

In summary, the time complexity of the above attack task is

$$Ŧ = (n + |\Lambda|) L_M[1/3, 1.923] + (n^4 |\Lambda|^4) + (n^4 |\Lambda|^4)(2|\Omega||\Lambda|)n$$
$$\approx 2n^5 |\Omega||\Lambda|^5.$$

Concretely speaking,

For $m = n = 80$ with $|\Lambda| = 2^{10}$ & $|\Omega| = 80$, $Ŧ > 2(2^{6.3})^5(2^{6.3})(2^{10})^5 = 2^{88} > 2^m$.

For $m = n = 96$ with $|\Lambda| = 2^{12}$ & $|\Omega| = 96$, $Ŧ > 2(2^{6.5})^5(2^{6.5})(2^{12})^5 = 2^{100} > 2^m$.

For $m = n = 112$ with $|\Lambda| = 2^{14}$ & $|\Omega| = 112$, $Ŧ > 2(2^{6.8})^5(2^{6.8})(2^{14})^5 = 2^{112} = 2^m$.

For $m = n = 128$ with $|\Lambda| = 2^{16}$ & $|\Omega| = 2^{12}$, $Ŧ > 2(2^7)^5(2^{12})(2^{16})^5 = 2^{128} = 2^m$.

For $m = n = 232$ with $|\Lambda| = 2^{32}$ & $|\Omega| = 2^{32}$, $Ŧ > 2(2^{7.8})^5(2^{32})(2^{32})^5 = 2^{232} = 2^m$.

Thus, the time complexity of the attack by presupposing $\ell(x_1) + \ell(x_2) = \ell(y_1) + \ell(y_2)$ is not less than $O(2^m)$ when $|\Lambda|$ and $|\Omega|$ are chosen suitably.

### 4.1.2 Ineffectualness of Guessing $\|W\|$

Owing to $80 \le \lceil \lg M \rceil \le 232$, $\bar{M}$ can be factorized in tolerable subexponential time, and further a value of $\|W\|$ can be guessed.

Adversaries may try to eliminate $W$ through $W^{\|W\|} \equiv 1$ (% $M$).

Raising either side of every equation $C_i \equiv (A_i W^{\ell(i)})^\delta$ (% $M$) to the $\|W\|$-th power yields

$$C_i^{\|W\|} \equiv (A_i)^{\delta\,\|W\|} \ \%\ M.$$

Suppose that the value of every $A_i \in \Lambda = \{2, 3, \ldots, Ƥ \mid 10 \le \lceil \lg Ƥ \rceil \le 32\}$ is guessed, or the possible values of every $A_i$ are traversed.

Let $C_i \equiv g^{u_i}$ (% $M$), and $A_i \equiv g^{v_i}$ (% $M$), where $g$ is a generator of $(\mathbb{Z}_M^*, \cdot)$. Then

$$u_i\|W\| \equiv v_i\|W\|\delta \ (\%\ \bar{M}) \ (i = 1, \ldots, n).$$

Notice that $u_i \ne v_i\delta$ (% $\bar{M}$), and $\{v_1, \ldots, v_n\}$ is not a super increasing sequence.

The above congruence is seemingly the MH transform [17]. Actually, $\{v_1\|W\|, \ldots, v_n\|W\|\}$ is not a super increasing sequence, and moreover there is not necessarily $\lceil \lg(u_i\|W\|) \rceil = \lceil \lg \bar{M} \rceil$.

Because $v_i\|W\| \in [1, \bar{M}]$ is stochastic, the inverse $\delta^{-1}$ % $\bar{M}$ not need be close to the minimum

$$\bar{M} / (u_i\|W\|),\ 2\bar{M} / (u_i\|W\|),\ \ldots,\ \text{or}\ (u_i\|W\| - 1)\bar{M} / (u_i\|W\|).$$

Namely $\delta^{-1}$ may lie at any integral position of the interval

$$[k\bar{M} / (u_i\|W\|),\ (k + 1)\bar{M} / (u_i\|W\|)],$$

where $k = 0, 1, \ldots, u_i\|W\| - 1$, which illustrates that the accumulation points of minima do not exist. Further observing, in this case, when $i$ traverses the interval $[2, n]$, the number of intersections of the intervals containing $\delta^{-1}$ is likely the max of $\{u_1\|W\|, \ldots, u_n\|W\|\}$ which is promisingly close to $\bar{M}$. Therefore, the Shamir attack by the accumulation point of minima is fully ineffectual [18].

Even if find out $\delta^{-1}$ through the Shamir attack method, because each of $\{v_1, \ldots, v_n\}$ has $\|W\|$ solutions, the number of potential sequences $\{g^{v_1}, \ldots, g^{v_n}\}$ is up to $\|W\|^n$.

Due to needing to verify whether $\{g^{v_1}, \ldots, g^{v_n}\}$ is a coprime sequence for each different sequence $\{v_1, \ldots, v_n\}$, the number of possible coprime sequences is in proportion to $\|W\|^n$. Hence, the initial $\{A_1, \ldots, A_n\}$ cannot be determined in subexponential time. Further, the value of $W$ cannot be computed, and the values of $\|W\|$ and $\delta^{-1}$ cannot be verified, which indicates that the MPP can also be resistant to the Shamir attack by the accumulation point of minima.

Additionally, the adversaries may divine the value of $A_i$ in about $O(|\Lambda|)$ time with $i \in [1, n]$, and compute $\delta$ by $v_i\|W\| \equiv u_i\|W\|\delta$ (% $\bar{M}$). However, because of $\|W\| \mid \bar{M}$, the equation will have $\|W\|$ solutions. Therefore, the time complexity of finding the original $\delta$ is at least

$$\begin{aligned} Ŧ &= (n + |\Lambda|)L_M[1/3, 1.923] + |\Lambda|\|W\| \\ &\ge (n + |\Lambda|)L_M[1/3, 1.923] + 2^{\lceil \lg Ƥ \rceil} 2^{m - \lceil \lg Ƥ \rceil} \\ &\ge 2^m. \end{aligned}$$

It is also not less than $O(2^m)$.

## 4.2 Security of the Compression Algorithm

The compression algorithm of which the input message is treated as only a block is the main body of the new non-iterative hash function, and thus, through it the four natural properties of the new hash function are embodied dominantly.

Clearly, the security of the compression algorithm depends on the security of the ASPP $\underline{d} \equiv \prod_{i=1}^n C_i^{\bar{b}_i}$ (% $M$), where $\bar{b}_i = \underset{.}{b}_i 2^{\partial_i}$ with $\partial_i = b_{i + (-1)^{\lfloor 2(i-1)/n \rfloor}(n/2)}$ and $\underset{.}{b}_i$ being a bit shadow.

In [6], we analyze the security of the ASPP $\bar{G} \equiv \prod_{i=1}^n C_i^{\underset{.}{b}_i}$ (% $M$) from the three aspects, discover no subexponential time solution to it, and contrarily, find some evidence which inclines people to believe that $\bar{G} \equiv \prod_{i=1}^n C_i^{\underset{.}{b}_i}$ (% $M$) is computationally harder than the DLP. Due to $\bar{b}_i = \underset{.}{b}_i 2^{\partial_i} \ge \underset{.}{b}_i$, the security conclusion about $\bar{G} \equiv \prod_{i=1}^{n} C_i^{\underset{\cdot}{b}_i}$ (% $M$) is also suitable for $\underline{d} \equiv \prod_{i=1}^{n} C_i^{\bar{b}_i}$ (% $M$) which is just another form of the ASPP. Hence $\underline{d} \equiv \prod_{i=1}^{n} C_i^{\bar{b}_i}$ (% $M$) has no subexponential time solution at present.

In what follows, we will analyze whether the compression formula $\underline{d} \equiv \prod_{i=1}^{n} C_i^{\bar{b}_i}$ (% $M$) satisfies the four natural properties of a hash function, and especially resists the three classical attacks or not.

In terms of Section 3.2, given the initial value ($\{C_i\}$, $M$) and a short message $b_1 \ldots b_n$, it is transparently easy to calculate the digest $\underline{d} \equiv \prod_{i=1}^{n} C_i^{\bar{b}_i}$ (% $M$).

### 4.2.1 Compression Algorithm Is Computationally One-way

Let $C_1 \equiv g^{u_1}$ (% $M$), …, $C_n \equiv g^{u_n}$ (% $M$), $\underline{d} \equiv g^{v}$ (% $M$), where $g$ is a generator of the group ($\mathbb{Z}_M^*$, $\cdot$), and easily found when $\lceil \lg M \rceil < 1024$.

Then, solving $\underline{d} \equiv \prod_{i=1}^{n} C_i^{\bar{b}_i}$ (% $M$) for $\bar{b}_1 \ldots \bar{b}_n$, namely $b_1 \ldots b_n$, is equivalent to solving

$$\bar{b}_1 u_1 + \ldots + \bar{b}_n u_n \equiv v \ (\% \ \bar{M}),$$

which is called an anomalous subset sum problem, shortly ASSP [6], and computationally at least equivalent to a subset sum problem (SSP) due to $\bar{b}_i = \underset{\cdot}{b}_i 2^{\partial_i} \geq \underset{\cdot}{b}_i \geq b_i \in [0, 1]$.

The SSP has been proved to be NP-complete in its feasibility recognition form [19], and its computational version, especially the density-high or length-big one, is NP-hard [1][20]. Hence, solving ASSP is at least NP-hard.

Moreover in the non-iterative hash function, there is $n \geq m = \lceil \lg M \rceil$ and $n \geq \bar{b}_i \geq b_i \in [0, 1]$. The knapsack density relevant to the ASSP $\bar{b}_1 u_1 + \ldots + \bar{b}_n u_n \equiv v$ (% $\bar{M}$) roughly equals

$$\begin{aligned} D &= \textstyle\sum_{i=1}^{n} \lceil \lg n \rceil / \lceil \lg M \rceil \\ &= n \lceil \lg n \rceil / m \\ &> \lceil \lg n \rceil \\ &> 1, \end{aligned}$$

which means that there exists many solutions to $\bar{b}_1 u_1 + \ldots + \bar{b}_n u_n \equiv v$ (% $\bar{M}$), namely the original solution cannot be determined, or will not occur in a reduced lattice base defined by LLL [21]. Notice that only such a $\langle \bar{b}_1, \ldots, \bar{b}_n \rangle$ from which a right bit string can be deduced will be a reasonable solution vector. Experiments show that when $D > 1$, the probability that the original solution or a reasonable solution is found through LLL lattice base reduction is almost zero [22].

Hence, LLL lattice base reduction attack on ASSP [21][23] is utterly ineffectual, which illustrates that even although a DLP with the modulus bit-length less than 1024 can be solved, the original or a reasonable $\bar{b}_1 \ldots \bar{b}_n$ cannot be found yet in DLP subexponential time, namely $\underline{d} \equiv \prod_{i=1}^{n} C_i^{\bar{b}_i}$ (% $M$) is computationally one-way.

### 4.2.2 Compression Algorithm Is Weakly Collision-free

Assume that $b_1 \ldots b_n \neq 0$ is a short message or a message digest from a classical hash function. By Definition 3, we easily understand that $\bar{b}_i = \underset{\cdot}{b}_i 2^{\partial_i} \leq n \ \forall i \in [1, n]$.

Given a short message $b_1 \ldots b_n \neq 0$, and let $b'_1 \ldots b'_n \neq 0$ be another short message to need to be found.

Let $\bar{b}_1 \ldots \bar{b}_n$ be the bit long-shadow string of $b_1 \ldots b_n$, and $\bar{b}'_1 \ldots \bar{b}'_n$ be the bit long-shadow string of $b'_1 \ldots b'_n$.

Let $l\hat{h}$ be the compression algorithm of the new non-iterative hash function described in Section 3.2. Hence, we have

$$\underline{d} = l\hat{h}(b_1 \ldots b_n) = \prod_{i=1}^{n} C_i^{\bar{b}_i} \ \% \ M,$$

and

$$\underline{d}' = l\hat{h}(b'_1 \ldots b'_n) = \prod_{i=1}^{n} C_i^{\bar{b}'_i} \ \% \ M,$$

where $\bar{b}_i = \underset{\cdot}{b}_i 2^{\partial_i}$ with $\partial_i = b_{i + (-1)^{\lfloor 2(i-1)/n \rfloor}(n/2)}$, and $\bar{b}'_i = \underset{\cdot}{b}'_i 2^{\partial'_i}$ with $\partial'_i = b'_{i + (-1)^{\lfloor 2(i-1)/n \rfloor}(n/2)}$.

If $\underline{d} = \underline{d}'$, there is

$$\prod_{i=1}^{n} C_i^{\bar{b}_i} \equiv \prod_{i=1}^{n} C_i^{\bar{b}'_i} \ (\% \ M).$$

Observe an extreme case.

Assume that $C_1 = \ldots = C_n = C$.

Owing to the max of $0 \leq \bar{b}_i \leq n$, we define logically

$$\prod_{i=1}^{n} C^{\bar{b}_i} \equiv \prod_{i=1}^{n} C^{(n+1)^{n-i} \bar{b}_i} \ (\% \ M).$$

Under the circumstances, if $\underline{d} = \underline{d}'$, then there is

$$\prod_{i=1}^{n} C^{(n+1)^{n-i}\bar{b}_i} \equiv \prod_{i=1}^{n} C^{(n+1)^{n-i}\bar{b}'_i}\ (\%\ M),$$

namely

$$C^{\sum_{i=1}^{n}(n+1)^{n-i}\bar{b}_i} \equiv C^{\sum_{i=1}^{n}(n+1)^{n-i}\bar{b}'_i}\ (\%\ M).$$

Let $z \equiv \sum_{i=1}^{n} \bar{b}_i (n+1)^{n-i}$ ($\%\ \bar{M}$), and $z' \equiv \sum_{i=1}^{n} \bar{b}'_i (n+1)^{n-i}$ ($\%\ \bar{M}$).

Correspondingly,

$$C^{z} \equiv C^{z'}\ (\%\ M).$$

We need to solve the above equation for $z'$.

If the order $\|C\|$ is known, let $z' = z + k\|C\|$, where $k \geq 1$ is an integer. Once a fit $k$ is found, there will be $C^{z} \equiv C^{z'}$ ($\%\ M$), and a bit string can be inferred from $\bar{b}'_1 \ldots \bar{b}'_n$. However, seeking $\|C\|$ is of the integer factorization problem (IFP) at present because the prime factors of $\bar{M}$ must be known.

In practice, $C_1, \ldots, C_n$ that are produced through the algorithm in Section 3.1 are pairwise unequal, which implies that for any given short message $b_1 \ldots b_n$, seeking another short message $b'_1 \ldots b'_n$ such that $\prod_{i=1}^{n} C_i^{\bar{b}_i} \equiv \prod_{i=1}^{n} C_i^{\bar{b}'_i}$ ($\%\ M$) is harder than the IFP in computational complexity, namely $b'_1 \ldots b'_n$ for $l\hat{h}(b_1 \ldots b_n) = l\hat{h}(b'_1 \ldots b'_n)$ cannot be found in IFP subexponential time.

Therefore, we say that the new non-iterative hash function is weakly collision-free.

#### 4.2.3 Compression Algorithm Is Resistant to Birthday Attack

First, observe an example of whether any two students in a class have the same birthday.

Suppose that the class has 23 students. If a teacher specifies a day (say February 12), then the chance that at least one student is born on that day is $(1 - (364 / 365)^{23}) \approx 6.11$ %. However, the probability that at least one student has the same birthday as any other student is around $(1 - (365 \times \ldots \times 343 / 365^{23})) \approx 50.73$ %, which prompts birthday attack on hash functions. Notice that the number $x$ of students will need increasing to 249 (> 365 / 2) if the teacher wants to make $(1 - (364 / 365)^{x}) = 50$ %.

Birthday attack, a type of strongly collision-free attack, is widely exploited for finding any two messages $\underline{m}$ and $\underline{m}'$ such that $\hat{h}(\underline{m}) = \hat{h}(\underline{m}')$, namely $(\underline{m}, \underline{m}')$ is a collision, where $\hat{h}$ is a hash function [24]. If the bit-length of a message digest is $m$, an adversary can find a collision $(\underline{m}, \underline{m}')$ such that $\hat{h}(\underline{m}) = \hat{h}(\underline{m}')$ with probability 50% in roughly $1.1774 \times 2^{m/2}$ time, namely with input of $1.1774 \times 2^{m/2}$ random messages [25].

However, to the new non-iterative hash, a collision is transformed into a mapping which is a type of weakly collision-free attack.

***Theorem 1:*** The new non-iterative hash function is resistant to birthday attack on the assumption that the MPP and ASPP have only exponential time solutions.

*Proof:*

Let $b_1 \ldots b_n$ and $b'_1 \ldots b'_n$ be two arbitrary different short messages, and $\bar{b}_1 \ldots \bar{b}_n$ and $\bar{b}'_1 \ldots \bar{b}'_n$ be their bit long-shadow strings respectively.

Suppose that $\underline{d} = \underline{d}'$, namely $\prod_{i=1}^{n} C_i^{\bar{b}_i} \equiv \prod_{i=1}^{n} C_i^{\bar{b}'_i}$ ($\%\ M$).

Because the ASPP has only exponential time solutions, we cannot directly solve $\underline{d} \equiv \prod_{i=1}^{n} C_i^{\bar{b}'_i}$ ($\%\ M$) for $\bar{b}'_1 \ldots \bar{b}'_n$.

In terms of the supposition, there is

$$\prod_{i=1}^{n} (A_i W^{\ell(i)})^{\delta \bar{b}_i} \equiv \prod_{i=1}^{n} (A_i W^{\ell(i)})^{\delta \bar{b}'_i}\ (\%\ M).$$

Further,

$$W^{\underline{k}\delta} \prod_{i=1}^{n} (A_i)^{\delta \bar{b}_i} \equiv W^{\underline{k}'\delta} \prod_{i=1}^{n} (A_i)^{\delta \bar{b}'_i}\ (\%\ M),$$

where $\underline{k} = \sum_{i=1}^{n} \bar{b}_i \ell(i)$, $\underline{k}' = \sum_{i=1}^{n} \bar{b}'_i \ell(i)\ \%\ \bar{M}$, and $\underline{k} - \underline{k}' < 4n(2\tilde{n} + 3)$.

Raising either side of the above congruence to the $\delta^{-1}$-th power yields

$$W^{\underline{k}} \prod_{i=1}^{n} A_i^{\bar{b}_i} \equiv W^{\underline{k}'} \prod_{i=1}^{n} A_i^{\bar{b}'_i}\ (\%\ M).$$

Without loss of generality, let $\underline{k} \geq \underline{k}'$. Because $(\mathbb{Z}_M^{*}, \cdot)$ is an Abelian group, we have

$$W^{\underline{k} - \underline{k}'} \equiv \prod_{i=1}^{n} A_i^{\bar{b}'_i} (\prod_{i=1}^{n} A_i^{\bar{b}_i})^{-1}\ (\%\ M).$$

Due to either $\bar{M} / 2$ = a prime or the least prime factor of $\bar{M} / 2 > 4n(2\tilde{n} + 3)$, there is

$$W^{2^k} \equiv (\prod_{i=1}^{n} A_i^{\bar{b}'_i - \bar{b}_i})^{((\underline{k} - \underline{k}') / 2^k)^{-1}}\ (\%\ M), \tag{1}$$

where $k \in [0, 46)$ is a small integer, $(\underline{k} - \underline{k}') / 2^k$ is a prime, and $W \in (1, \bar{M})$ as a component of a private key is determinate, which manifests that if $\bar{b}_1 \ldots \bar{b}_n$ and $\bar{b}'_1 \ldots \bar{b}'_n$ satisfy (1), there will be $\underline{d} = \underline{d}'$.

For clear explanation, (1) is written as the form of a function:

$$x^{2^k} \equiv (\textstyle\prod_{i=1}^{n} A_i^{\bar{b}'_i - \bar{b}_i})^{((\underline{k} - \underline{k}') / 2^k)^{-1}} \ (\% \ M). \tag{2}$$

Since $\bar{M}$ contains only one 2-factor, (2) has only two solutions when $k \neq 0$.

In other words, we may define a mapping from $\{0, 1\}^n \times \{0, 1\}^n$ to $\{1, \ldots, \bar{M}\}$:

$$\Psi(b_1 \ldots b_n, b'_1 \ldots b'_n) \equiv (\textstyle\prod_{i=1}^{n} A_i^{\bar{b}'_i - \bar{b}_i})^{((\underline{k} - \underline{k}') / 2^k)^{-1}} \ (\% \ M),$$

where $\bar{b}_i = \underset{\cdot}{b}_i 2^{\partial_i}$, $\bar{b}'_i = \underset{\cdot}{b}'_i 2^{\partial'_i}$, $\underline{k} = \sum_{i=1}^{n} \bar{b}_i \ell(i)$, $\underline{k}' = \sum_{i=1}^{n} \bar{b}'_i \ell(i) \ \% \ \bar{M}$, $k \in [0, 46)$ is a integer, and $(\underline{k} - \underline{k}') / 2^k$ is a prime.

Therefore, only if $\Psi(b_1 \ldots b_n, b'_1 \ldots b'_n) = W^{2^k}$ with $k \in [0, 46)$, can there exists $\underline{d} = \underline{d}'$. Obviously, $\forall$ $(b_1 \ldots b_n, b'_1 \ldots b'_n) \in \{0, 1\}^n \times \{0, 1\}^n$, the probability that $\Psi(b_1 \ldots b_n, b'_1 \ldots b'_n) = W^{2^k}$ is nearly $k/2^m$ (the number of values in the form of $W^{2^k}$ is at most $k$).

Further, let $\underline{n}$ be the number of $(b_1 \ldots b_n, b'_1 \ldots b'_n)$'s which need to be inputted in order to find at least one $(b_1 \ldots b_n, b'_1 \ldots b'_n)$ such that $\Psi(b_1 \ldots b_n, b'_1 \ldots b'_n) = W^{2^k}$ with probability 50%, namely to find any two messages $b_1 \ldots b_n$ and $b'_1 \ldots b'_n$ such that $\hat{lh}(b_1 \ldots b_n) = \hat{lh}(b'_1 \ldots b'_n)$ with probability 50%. Then, $\underline{n}$ satisfies $1 - ((2^m - k)/2^m)^{\underline{n}} = 50\%$. Resorting to computation, we see that $\underline{n}$ is nearly equal to $2^{m-1}$ with $k \in [0, 46)$.

The $2^{m-1}$ is far larger than the threshold $1.1774 \times 2^{m/2}$ for the effective birthday attack. The reason is that a hidden restriction is imposed on the input $(b_1 \ldots b_n, b'_1 \ldots b'_n)$, which is easily understood as the number of students of the class needs to be increased for finding any two students who have both the same birthday and the same *gender* with probability 50%.

Additionally, because a private key $(\{A_i\}, \{\ell(i)\}, W, \delta)$ is unknown for the adversary, and the MPP is intractable, it is also infeasible that the adversary finds specific $b_1 \ldots b_n$ and $b'_1 \ldots b'_n$ which make (1) hold according to the private key.

Therefore, the new non-iterative hash can be resistant to the birthday attack, and at present, its security is nearly the $O(2^m)$ magnitude, but not $O(2^{m/2})$. □

### 4.2.4 Compression Algorithm Is Resistant to Meet-in-the-middle Attack

Meet-in-the-middle dichotomy used for attack on an intended expansion of a block cipher was first developed by Diffie and Hellman in 1977 [26]. Section 3.10 of [1] brings forth a meet-in-the-middle attack algorithm for solving a subset sum problem.

Let $b_1 \ldots b_n$ be a short message, and its digest be $\underline{d} \equiv \prod_{i=1}^{n} C_i^{\bar{b}_i} \ (\% \ M)$.

If $b_{n/2} = b_n = 1$ (thus, any bit *shadow* on the left of the middle point has no relation with bits on the right), an adversary may attempt to attack the ASPP $\underline{d} \equiv \prod_{i=1}^{n} C_i^{\bar{b}_i} \ (\% \ M)$ by the meet-in-the-middle method.

However, owing to $\bar{b}_i = \underset{\cdot}{b}_i 2^{\partial_i}$ with $\partial_i = b_{i + (-1)^{\lfloor 2(i-1)/n \rfloor}(n/2)}$ for every $i \in [1, n]$, when $i$ is from 1 to $n / 2$, there exists

$$\bar{b}_1 \ldots \bar{b}_{n/2} = (\underset{\cdot}{b}_1 2^{b_{1+n/2}}) \ldots (\underset{\cdot}{b}_{n/2} 2^{b_n}),$$

which involves all the bits of the short message, namely a reasonable middle point does not exist.

If a fork is selected in proportion to $(n / 3 : 2n / 3)$ or $(n / 4 : 3n / 4)$, the right of the fork substantially still involves all the bits $b_1, \ldots, b_n$.

For instance, let $n = 12$, a short message (a bit string) $= b_1 \ldots b_{12}$, and a fork be to $(4 : 8)$, then

$$\bar{b}_5 \ldots \bar{b}_{12} = (\underset{\cdot}{b}_5 2^{b_{11}})(\underset{\cdot}{b}_6 2^{b_{12}})(\underset{\cdot}{b}_7 2^{b_1})(\underset{\cdot}{b}_8 2^{b_2})(\underset{\cdot}{b}_9 2^{b_3})(\underset{\cdot}{b}_{10} 2^{b_4})(\underset{\cdot}{b}_{11} 2^{b_5})(\underset{\cdot}{b}_{12} 2^{b_6})$$

involves all the bits $b_1, \ldots, b_{12}$.

The above dissection manifests that the meet-in-the-middle attack is essentially ineffectual on the new non-iterative hash function. Therefore, even if $n = m$, namely the input length = the output length of the function, the time complexity of the attack task is still $O(2^m)$ at present, but not $O(m 2^{m/2})$.

Besides, unlike $\sum_{i=1}^{n} c_i = \sum_{i=1}^{n} b_i c_i + \sum_{i=1}^{n} \neg b_i c_i$ in the SSP, there is not

$$\textstyle\prod_{i=1}^{n} C_i = \prod_{i=1}^{n} C_i^{\bar{b}_i} \prod_{i=1}^{n} C_i^{\neg \bar{b}_i} \ (\% \ M)$$

in the ASPP, where $\neg \bar{b}_i$ is the bit long-shadow of $\neg b_i$, which implies there does not exist an easy relation between the ASPP $\underline{d} \equiv \prod_{i=1}^{n} C_i^{\bar{b}_i} \ (\% \ M)$ and the dichotomy.

### 4.2.5 Compression Algorithm Is Resistant to Multi-block Differential Attack

The [27] and [28] show that multi-block near differential attack is effective on the iterative hash functions MD5, SHA-0, SHA-1, and SHA-256 which have multiple block-inputs and the Merkle-Damgård structure [4][5].

It is well known that MD5, SHA-0, or SHA-1 will execute a number of rounds of inner manipulation for every input block, and each round of the inner manipulation consists of linear arithmetics and/or logic operators such as *addition*, *shift*, *and*, *not*, *exclusive or*, etc.

The input of the new non-iterative hash function is a short message which may be treated as only one block. Its inner manipulation consists of at most $2n$ modular multiplications which is nonlinear and intricate, which indicates that the differential analysis of $\bar{g} \equiv \prod_{i=1}^{n} C_i^{\bar{b}_i}$ (% $M$) loses a basis.

Furthermore, in the new non-iterative hash, the inner nonlinear manipulation leads to the fierce snowslide effect and strong noninvertibility (see Section 4.2.1), and makes it impossible to derive a set of sufficient conditions which ensure that the collision differential characteristics hold for two short messages which are expected to produce a collision.

Therefore, the new non-iterative hash is substantially distinct from the classical iterative hashes MD5, SHA-0, SHA-1 etc, and the multi-block near differential attack suitable for the classical iterative hashes will be utterly ineffective on the new non-iterative hash function.

### 4.2.6 Compression Algorithm Is Strongly Collision-free

Firstly, it is known from Section 4.2.2 that the new non-iterative hash function $l\hat{h}$ is weakly collision-free.

Secondly, for any arbitrary short message $b_1 \ldots b_n$, if want to find another short message $b'_1 \ldots b'_n$ such that $l\hat{h}(b_1 \ldots b_n) = l\hat{h}(b'_1 \ldots b'_n)$, adversaries must take $\bar{b}'_1 \ldots \bar{b}'_n$ from $\prod_{i=1}^{n} C_i^{\bar{b}_i} \equiv \prod_{i=1}^{n} C_i^{\bar{b}'_i}$ (% $M$), and further acquire the bit string $b'_1 \ldots b'_n$. It is known from Section 4.2.2 that such a collision problem is computationally harder than IFP now.

Thirdly, the new non-MDS hash is resistant to classical or efficient attacks in common use — the birthday attack, meet-in-the-middle attack, and multi-block differential attack for example.

Lastly, any subexponential time algorithm for solving the ASPP $\bar{g} \equiv \prod_{i=1}^{n} C_i^{\bar{b}_i}$ (% $M$) is not found yet [29], and the most efficient method of solving $\bar{g} \equiv \prod_{i=1}^{n} C_i^{\bar{b}_i}$ (% $M$) is brute force attack so far. The analysis manifests that the security of the new non-iterative hash gets the $O(2^m)$ magnitude at present.

In sum, the new hash function is strongly collision-free. Further, we may give a related theorem.

***Theorem 2:*** If any arbitrary collision of the new non-iterative hash function can be found in subexponential time, the ASPP $\prod_{i=1}^{n} C_i^{\bar{y}_i} \equiv 1$ (% $M$) can be solved in subexponential time, where $\bar{y}_i \in [-n, n]$ is the difference of two bit long-shadows at the same position.

*Proof:*

According to Definition 3, it is easy to understand that for each $\bar{b}_i$, there is $0 \le \bar{b}_i \le n$.

Let $b_1 \ldots b_n \ne b'_1 \ldots b'_n \ne 0$ be two arbitrary bit strings, $\bar{b}_1 \ldots \bar{b}_n$ and $\bar{b}'_1 \ldots \bar{b}'_n$ be respectively two corresponding bit long-shadow strings.

Again let $\bar{y}_i = \bar{b}_i - \bar{b}'_i$, and then there is $\bar{y}_i \in [-n, n]$.

Since the interval $[-n, n]$ is wider than $[0, n]$, similar to $\bar{g} \equiv \prod_{i=1}^{n} C_i^{\bar{b}_i}$ (% $M$), the ASPP $\prod_{i=1}^{n} C_i^{\bar{y}_i} \equiv 1$ (% $M$) with $\bar{y}_i \in [-n, n]$ has no subexponential time solution [29], and is only faced with brute force attack.

Assume that $\prod_{i=1}^{n} C_i^{\bar{b}_i} \equiv \prod_{i=1}^{n} C_i^{\bar{b}'_i}$ (% $M$) is a found collision between two arbitrary bit strings $b_1 \ldots b_n$ and $b'_1 \ldots b'_n$ in subexponential time.

From $\prod_{i=1}^{n} C_i^{\bar{b}_i} \equiv \prod_{i=1}^{n} C_i^{\bar{b}'_i}$ (% $M$), we have

$$\prod_{i=1}^{n} C_i^{\bar{b}_i - \bar{b}'_i} \equiv 1 \text{ (\% } M\text{)}.$$

Let $\bar{y}_i \equiv \bar{b}_i - \bar{b}'_i \in [-n, n]$, and then

$$\prod_{i=1}^{n} C_i^{\bar{y}_i} \equiv 1 \text{ (\% } M\text{)},$$

which means that the ASPP $\prod_{i=1}^{n} C_i^{\bar{y}_i} \equiv 1$ (% $M$) can be solved efficiently in subexponential time. It is in direct contradiction to the fact.

Therefore, the new non-iterative hash function is strongly collision-free. □

## 5 Comparison with the Chaum-Heijst-Pfitzmann Hash

The Chaum-Heijst-Pfitzmann hash function is provably secure, and defined as follows [12]:

$$\hat{h}: w_1, w_2 \mapsto \hat{h}(w_1, w_2) = \alpha^{w_1}\beta^{w_2} \,\%\, p \quad (\{0, ..., q-1\}^2 \to \mathbb{Z}_p - \{0\}),$$

where $w_1$ and $w_2$ are the two complementary parts of a short message, $p$ and $q$ ($= (p-1)/2$) are two big primes, and $\alpha$ and $\beta$ are two generators of the group $(\mathbb{Z}_p^*, \cdot)$.

Hence, the Chaum-Heijst-Pfitzmann hash function based on the difficulty of the DLP $\beta = \alpha^x \,\%\, p$ compresses a short message of $2(\lceil \lg p \rceil - 1)$ bits into a digest of $\lceil \lg p \rceil$ bits.

Let $\lceil \lg p \rceil = 1024$, and then the time complexity of computing $\log_\alpha \beta \,\%\, p$ is $2^{80}$ according to the subexponential time $L_p[1/3, 1.923]$ [1], which means that the security of the Chaum-Heijst-Pfitzmann hash is the $2^{80}$ magnitude when $\lceil \lg p \rceil = 1024$.

Let $\lceil \lg M \rceil = 80$, and then the time complexity of solving the ASPP $\underline{d} = \prod_{i=1}^{n} C_i^{\bar{b}_i} \,\%\, M$ for $\bar{b}_1, \ldots, \bar{b}_n$ is also $2^{80}$ since the ASPP only has an exponential time solution at present [29], which means that the security of the new non-iterative hash is also the $2^{80}$ magnitude when $\lceil \lg M \rceil = 80$. Besides, let the bit-length $n = 2046$ of a short message $(w_1, w_2) = (b_1 \ldots b_{1023}, b_{1024} \ldots b_{2046}) = b_1 \ldots b_n \neq 0$.

Under the same security, may draw a comparison between the new non-iterative hash (the JUNA hash) and the Chaum-Heijst-Pfitzmann hash.

**Table 1**. Comparison between two non-iterative hashes.

| | Chaum-Heijst-Pfitzmann hash | JUNA hash |
|---|---|---|
| Running time (bit operations) | $2(4\lceil \lg p \rceil^3) = 8589934592$ | $4nm^2 = 52428800$ |
| Compression rate | $1024 / 2046 \approx 50.05\%$ | $80 / 2046 \approx 3.91\%$ |
| Resistant to birthday attack | No<br>because the number of $(w_1, w_2)$'s needed during birthday attack is about $2^{\lceil \lg p \rceil / 2} = 2^{512}$, and larger than $2^{80}$ which is the security magnitude of the DLP. | Yes<br>because the number of $b_1 \ldots b_n$'s needed during birthday attack is about $2^{\lceil \lg M \rceil / 2} = 2^{40}$, and smaller than $2^{80}$ which is the security magnitude of the ASPP. |
| Provably strongly collision-free | Yes<br>on the assumption that a DLP has a subexponential time solution. | Yes<br>on the assumption that an ASPP has an exponential time solution. |

In summary, the JUNA hash has some advantages over the Chaum-Heijst- Pfitzmann one, and relatively the JUNA hash may be regarded lightweight.

## 6 Reformation of a Classical Hash Function

Because the new non-iterative hash function is resistant to birthday attack and meet-in-the-middle attack, a classical hash function of which the output is $m$ bits, and the security is intended to be the $O(2^{m/2})$ magnitude may be reformed into a compact hash function of which the output is $m/2$ bits, and the security is still equivalent to the $O(2^{m/2})$ magnitude [30].

For example, let $b_1 \ldots b_{128}$ be the output of MD5 [31], $\underline{b}_1 \ldots \underline{b}_{128}$ be its bit long-shadow string, and $\lceil \lg M \rceil = 64$. Then, regard $\underline{d} = \prod_{i=1}^{128} C_i^{\underline{b}_i} \,\%\, M$ as the 64-bit output of the reformed MD5 with the equivalent security, where $C_i = (A_i W^{\ell(i)})^\delta \,\%\, M$ which is produced by the algorithm in Section 3.1.

Again for example, let $b_1 \ldots b_{160}$ be the output of SHA-1[1], $\underline{b}_1 \ldots \underline{b}_{160}$ be its bit long-shadow string, and $\lceil \lg M \rceil = 80$. Then, regard $\underline{d} = \prod_{i=1}^{160} C_i^{\underline{b}_i} \,\%\, M$ as the 80-bit output of the reformed SHA-1 with the equivalent security.

The above two examples indicate that we may exchange time for space when the related security remains unchanged.

## 7 Conclusion

In the paper, the authors propose a new non-iterative hash function which contains the initialization algorithm and the compression algorithm, and converts a short message or a message digest of $n$ bits into a string of $m$ bits, where $80 \le m \le 232$ and $80 \le m \le n \le 4096$.

The authors analyze the security of the new non-iterative hash function. The analysis shows that the new non-iterative hash is computationally one-way, weakly collision-free, and strongly collision-free. Moreover, at present, any subexponential time algorithm for attacking the new non-iterative hash is not found, and its security is to the $O(2^m)$ magnitude.

Especially, the analysis illustrates that the new non-iterative hash function is resistant to birthday attack and meet-in-the-middle attack, and that the running time of its compression algorithm is $O(n m^2)$ bit operations.

The application of the new hash may be extended. In recent years, the ECC-160 digital signing scheme, an analogue of the ElGamal digital signing scheme based on the DLP in an elliptic curve group over a finite field [32][33], and some lightweight digital signing schemes — the optimized version of the REESSE1+ digital signing scheme [6] for example have been utilized for RF ID (Radio Frequency Identity) tags or non-RF ID tags [34][35][36]. While a RF ID tag contains an IC chip which is used to store signatures and other data, an non-RF ID tag, a BFID [37] — for example contains no IC chip because a signature from a lightweight or ultra-lightweight signing scheme may be symbolized in short length, and printed directly on a papery tag or label. At present, such tags are applied to the identification, authentication, or anti-forgery of financial-notes, certificates, diplomas, and commodities, particularly including food and drug.

Hence, the new non-iterative hash function opens a door to convenience for the utilization of a lightweight digital signing scheme of which the modulus length is not greater than 160 bits.

## Acknowledgment

The authors would like to thank the Academicians Jiren Cai, Zhongyi Zhou, Jianhua Zheng, Changxiang Shen, Zhengyao Wei, Binxing Fang, Guangnan Ni, Andrew C. Yao, Xicheng Lu, Wen Gao, Wenhua Ding, and Xiangke Liao for their important advice and helps.

The authors also would like to thank the Professors Dingyi Pei, Jie Wang, Ronald L. Rivest, Moti Yung, Adi Shamir, Dingzhu Du, Mulan Liu, Huanguo Zhang, Yixian Yang, Maozhi Xu, Hanliang Xu, Dengguo Feng, Xuejia Lai, Yongfei Han, Yupu Hu, Dongdai Lin, Rongquan Feng, Ping Luo, Jianfeng Ma, Lusheng Chen, Chuankun Wu, Lin You, Wenbao Han, Bogang Lin, Lequan Min, Qibin Zhai, Hong Zhu, Renji Tao, Zhiying Wang, Quanyuan Wu, and Zhichang Qi for their important suggestions and corrections.

## Appendix A: An Example

Let $\lceil \lg M \rceil = 80$, and $n = 256$.

**Solving the MPP:**

Given $M$ = 636743755563737235857207, and $\{C_1, \ldots, C_{256}\}$ =
{394375509141369037703184,554405328844801192217442,398990392120059456829699,636068710931207324336104,179366946033260810673265,182182128843950184496233,283653432762798960694200,391748237477785007893514,944612305738333399041634,146396573827145853058025,544816169334706503213027,364481169034548457969826,477943409648888873528887,495981229119127077122569,303247879531079652865837,302610401146719645640 35,604806200768061661948367,226709912769734878042146,211067870835444250207 47,450585510787322862879583,113889741803376766817431,337798241076366776900 00,624343348434427417711884,813943362892832145 4057,965063821903110576142 48,359344008158083077617116,475087369983772394584265,286675906747363274106643,273904561106043852824719,290154030115540709591119,542337668830272754302104,424209565234481301351243,482163813841492061131471,127934386844210811350835,594961208610220091706500,368457620191339441765069,333246120093389698485472,240036277940820391108175,326079559057243941942753,180855393210421934443585,558957548924545352698752,116963332670423702444319,620364395658763217288588,747080208426088619619 19,338603136005253750049019,618279924416273562129128,600081310839835683212541,606675873657517853028369,215973513658356020420635,539913213636759819602147,673973908015845784472 55,102206491211043454760486,171011183472338301996410,556402611627196680689898,381458105511009220697638,532956153792890202951438,360925851265173951197208,216608387452547613908 74,113278415082646883610336,587295387093175644250777,441835526319605486874262,495857237690484091878476,427476083339017325472093,414844423032073223749402,267957140905

582483315581,4077754020614154847965 91,4733298477518247965092 35,2377305409375710613 36
583,454275729099091444480453,250663187262216724468 27,213153434564424036920709,7695544
3512116632014080,577719850708310853721751,296881334499832564905758,28082635141801498
4314614,305079484542031100608532,369948879483802705833417,178519896368431501154183,15
594444390662190096 7508,358879495202308295530086,538801869715990957229057,46219020894
0699793771101,401751978131978482609 86,262448765064486865723793,22026207758871926 9492
112,192432627187402744418430,203874081871546080137836,273615761529636585860982,470964
1831576687520 2081,545718729741407541033298,256902461410255239515414,8679653331105043
1751282,615699406626702658312424,727769371460938593404 0,623661508518352474833795,3413
387510788374616962 60,833873585928670884916 34,331745118809598203756547,14600841305494
0870474217,377718668238650499325708,573308954069191320954876,192583455470829260572 52
6,257636756198775697553561,457854147247221048492853,295005661335709158380650,61310489
6771788170321637,476640631132253173570 72,112465310193651528643453,239327146015505183
869321,428852058761047961206417,621034609683055018803847,138845629932573936666694,389
98831706319699432871 0,625798568384070501018232,167048576453301484653376,639985062348
1354811793,253312083066930370988 2,441364010361767243247859,215298769730452968440469,7
888527600938564520565 6,366142537012652261414173,106705557479793492902577,34204768859
6789250089719,383295777538093497752089,226822823393548166858605,4547220097880346470 4
1861,964110073867307171558 15,152271197161087713633906,425287855627697178809174,226205
831082936831340019,791454916957158673564 27,243448386701422251112551,3465948018151363
7217315,627169519771260009749 93,469120356154738212445264,618660910804439681244744,484
254940080337537672234,572166973409032644768790,366057954716044986537 5,26312791843352
9780572115,170212898238335696139941,422732042511190107949564,308446040612533299953 10
5,373003147046146839017941,509025463714927591001093,375881626021462104944196,58745770
829970890902335 7,115257190305617586537407,610881911245478642078000,48375260940199943
3108445,217261946718280470713735,533424298980600127268003,361984585662190582028097,13
434806614175091250179 8,403240403838225119367554,313367491914963584952010,24943420419
8818855115174,539488866558263483937488,399519957905911405204918,49133357241379952290
6743,616764503083569121724952,498941513621940376156838,360115355217060253333938,28675
6596346655156944400,543341681019728138219968,240993764872128300299962,18798947385919
6573392152,137421203010702125156501,489873292467205032012327,61296148343986720122971
6,633009400619994839941913,442965146354422859554362,322638110572502910167370,32234558
3769379567431049,462590776934506038776857,368824221513851136474572,22379442394454434
9100743,442946162562545923022539,535412005420704431112529,434535990291959608671501,60
5645010994779584866952,807020629150144196515 4,493511370954416873059008,6188360274190
14613362898,590662580024211355162012,457494664211307406557064,963613477007484916633 8
4,120583811596327848299164,180442197235245703784100,405740657284513824054844,40431194
047718221412170,468082207913731037323835,229468643859253759600978,598297710404864974
354341,209048001585555967856547,457743106588718408708912,596519246673853139695397,608
540108389989364933186,555583430086257539238992,353434117833141924681370,382842801308
302520061705,492071882418698492159424,621445795157335823489745,25007642847726458168 5
569,546213632312565034207207,497298374430742379786584,191037533658442834834989,593133
366832103108156787,212457956727128031940975,620485991163132474252386,757713731242739
57235870,260871794980499581085477,549333245096281904234582,443239692067375141612071,5
515447797079994110767 56,288443772113295541911443,186925867422825217898560,3920573957
45465277837836,240883535976209539688869,549315739766192959945090,369022547903597352 5
30869,235207478202534037876752,119244538852522553537061,639453869674468969832 53,44799
303786915069584716 0,349184653845911760345919,410978297720843053424788,29876812535317
8719219809,237490662717517417924479,601270004230179754794434,340071233305985567657 21
9,554975899833724562810348,159174106445636336094312,694471509751687880939 06,318489470
752076358290636,569233492081487464852735,486228321190255110795019,584931011042787342
545814,278566431285608341099 8,144387067223408888572 34,220309245141837703800089,135194
41311645009571824 4,837465326571267492941 70,746889134285482770952 22,2372363655298962 9
8380585,148733606480086004988750,608490204061290555741 11,532867705593657608077 06,5505
268747743023456354 30,139918462219083995087941,328129290014413336506695,3975735392751
3730348711,119152179893933079618 56,343253875442491197058730,541569087399401325673659,

50037875839854944963 0036},

seek the original ($\{A_i\}$, $\{\ell(i)\}$, $W$, $\delta$) by $C_i \equiv (A_i W^{\ell(i)})^{\delta}$ (% $M$)($i$ = 1, …, 256), where $A_i \in \Lambda = \{2, 3, \ldots, 287117\}$, and $\ell(i) \in \Omega = \{+/-5, +/-7, \ldots, +/-515\}$.

**Solving the ASPP:**

Given the initial value ($M$, $\{C_1, \ldots, C_{256}\}$) that precedes, a short message $\underline{m} = \{b_1, \ldots, b_{256}\}$ =
{1,0,0,1,0,0,0,1,0,1,0,1,1,0,1,1,0,0,0,1,0,0,0,1,1,0,0,0,1,0,1,0,0,0,0,1,1,1,0,0,1,0,1,1,1,1,0,0,0,1,1,0,0,1,1
,0,0,0,0,0,1,1,1,1,1,1,1,0,0,1,1,1,0,0,1,0,1,0,1,0,0,1,0,0,1,0,1,1,1,1,0,0,1,1,1,1,1,0,1,1,0,0,1,1,0,0,1,1,1,1,
0,1,1,0,1,0,0,1,0,0,0,0,0,0,0,0,1,0,0,1,1,1,0,1,0,1,1,1,1,1,1,1,0,0,1,0,0,0,1,0,0,0,0,1,1,0,1,0,0,0,1,0,1,1,0,
0,0,1,1,1,0,1,1,0,0,1,1,0,0,0,1,1,0,0,0,0,0,0,0,0,0,0,1,0,1,0,0,0,1,1,0,0,0,1,0,0,0,0,0,0,0,0,0,0,1,1,1,1,1,1,0,
0,0,0,0,0,1,1,0,1,1,1,0,0,0,0,0,0,0,1,0,1,1,1,1,1,0,0,1,1,1,0,0,0,1,0,0},
and the digest $\underline{d}$ = 566936505785934227489970,
seek a collision with $\underline{m}$ by $\underline{d} \equiv \prod_{i=1}^{n} C_i^{\underline{b}_i}$ (% $M$), where $\underline{b}_i = \underset{\cdot}{b}_i 2^{\partial_i}$ with $\partial_i = b_{i+(-1)^{\lfloor 2(i-1)/n \rfloor}(n/2)}$.